\documentclass[amssymb,aps,prb,multicol,groupedaddress,showpacs,showkeys,floatfix]{revtex4}
\usepackage{amsmath}
\usepackage{epsfig}

\begin{document}

\title{
Spin-subband populations and spin polarization of quasi two-dimensional carriers \\
under in-plane magnetic field}

\author{Constantinos Simserides$^{1,2}$
\footnote{E-mail: {\sf simserides@upatras.gr},
URL: {\sf http://www.matersci.upatras.gr/simserides}
}
}

\affiliation{
$^1$University of Patras, Materials Science Department, Panepistimiopolis, Rio,
GR-26504, Patras, Greece \\
$^2$University of Athens, Physics Department, Panepistimiopolis, Zografos,
GR-15784, Athens, Greece}

\date{\today}

\begin{abstract}
Under an in-plane magnetic field,
the density of states of quasi two-dimensional carriers
deviates from the occasionally stereotypic step-like form
both quantitatively and qualitatively.
For the first time in the literature
as far as we know,
we study how this affects
the spin-subband populations and the spin-polarization
as functions of the temperature, $T$,
and the in-plane magnetic field, $B$,
for narrow to wide
dilute-magnetic-semiconductor quantum wells.
We examine a wide range of material and structural parameters,
focusing on the quantum well width,
the magnitude of the spin-spin exchange interaction,
and the sheet carrier concentration.
Generally, increasing $T$,
the carrier spin-splitting, $U_{o\sigma}$, decreases,
augmenting the influence of the ``minority''-spin carriers.
Increasing $B$, $U_{o\sigma}$ increases and accordingly
carriers populate ``majority''-spin subbands while they abandon
``minority''-spin subbands.
Furthermore, in line with the density of states modification,
all energetically higher subbands become gradually depopulated.
We also indicate the ranges where the system is completely spin-polarized.
\end{abstract}

% 85.75.-d Magnetoelectronics; spintronics: devices exploiting spin polarized
%          transport or integrated magnetic fields
% 75.75.+a Magnetic properties of nanostructures
% 75.50.Pp Magnetic semiconductors


\maketitle

%%%%%%%%%%%%%%%%%%%%%%%%%%%%%%%%%%%%%%%%%%%%%%%%%%%%%%%%%%%%%%%%%%%%%%%%%%%%%%%
\section{Preamble}%%%%%%%%%%%%%%%%%%%%%%%%%%%%%%%%%%%%%%%%%%%%%%%%%%%%%%%%%%%%%
\label{sec:preamble}%%%%%%%%%%%%%%%%%%%%%%%%%%%%%%%%%%%%%%%%%%%%%%%%%%%%%%%%%%%
%%%%%%%%%%%%%%%%%%%%%%%%%%%%%%%%%%%%%%%%%%%%%%%%%%%%%%%%%%%%%%%%%%%%%%%%%%%%%%%
An in-plane magnetic field
applied to a {\it quasi} two-dimensional system
distorts the equal-energy surfaces \cite{ees,sj}
or equivalently the density of states \cite{lyo:94,simserides:99} (DOS).
An interplay between spatial and magnetic confinement is established
and -properly- it is necessary to compute self-consistently
the energy dispersion, $E_{i,\sigma}(k_x)$,
where
$i$ is the subband index,
$\sigma$ denotes the spin, $k_x$ is the in-plane wave vector
perpendicular to the in-plane magnetic field, $B$ (applied along $y$),
and $z$ is the growth axis.
Hence, the envelope function along $z$ depends on $k_x$ i.e.,
$\psi_{i,\sigma,k_x,k_y}({\bf r})
\propto \zeta_{i,\sigma,k_x}(z) exp(i k_x x) exp(i k_y y)$.
This modification has been realized in
{\it magnetotransport} \cite{magnetotransport} and
{\it photoluminescence} \cite{PL} experiments.
An impressive fluctuation of the {\it in-plane magnetization}
in dilute-magnetic-semiconductor (DMS) structures
in cases of strong competition between spatial and magnetic confinement
has been predicted at low enough temperatures \cite{simserides-prb-2004}
and a compact DOS formula holding for any type of interplay
between spatial and magnetic confinement already exists \cite{simserides-prb-2004}.

Although this DOS modification can be extremely significant
both quantitatively and qualitatively
it is sometimes neglected without a second thought.
Naturally, in the limit of very narrow quantum wells (QWs) or for $B \to 0$,
the DOS preserves the ideal step-like form.
The ``opposite'' asymptotic limit is a simple saddle point,
where the DOS diverges logarithmically \cite{lyo:94}.
However, generally, the van Hove singularities which show up
are not simple saddle points \cite{simserides:99}.
Summarizing, models which ignore the above DOS modifications
can only be applied to very narrow QWs or for $B \to 0$.

During the last years,
the progress in growth, characterization and understanding of
transition-metal-doped semiconductors has been impressive
\cite{ohno,dietl-ohno,jungwirth_et_al_review}.
As a result, new phenomena have been discovered, e.g.
tunnel magnetoresistance,
spin-dependent scattering,
interlayer coupling due to carrier polarization,
electrical electron and hole spin injection,
and electric field control of ferromagnetism \cite{ohno,dietl-ohno}.
Usually the host material is a III-V semiconductor
\cite{ohno,dietl-ohno,jungwirth_et_al_review}.
For example, in (Ga,Mn)As or in (In,Mn)As,
Mn substitutes a small fraction of cations
providing holes and local magnetic moments.
Hence, the corresponding structures utilize the valence band.
The highest ferromagnetic transition temperature, $T_C$,
reported so far for III-V-based
valence-band magnetic semiconductors is
173 K in (Ga,Mn)As epilayers \cite{jungwirth_et_al_review}.

In II-VI materials, Mn provides only local magnetic moments.
The corresponding heterostructures,
for example ZnSe/Zn$_{1-x-y}$Cd$_x$Mn$_y$Se,
utilize either the conduction or the valence band,
depending on the type of dopants used in the barriers,
namely, donors (e.g. Cl, I) or acceptors (e.g. Li), respectively.
In the present article we investigate such a system where
either the conduction band or the valence band
can be exploited for spintronic applications.
The key material of each structure (e.g. ZnSe, CdTe etc) may possess
quite different material parameters e.g. positive or negative $g$ factors \cite{gstar}.
We also note that the band gap of common II-VI crystals
covers all the range from the ultraviolet to the infrared \cite{textbook}.
Interestingly the existence of ferromagnetic order
in n-doped (Cd,Mn)Te based structures
-at extremely low temperatures- has been suggested
both experimentally and theoretically
\cite{teran,koenig_macdonald}.

In a recent publication \cite{simserides-prb-2004}
we restricted ourselves to DMS structures utilizing the conduction band and
to very low temperatures.
We studied the spin-subband populations, the internal and free energy,
the Shannon entropy, and the in-plane magnetization $M$
as functions of the in-plane magnetic field, for different degrees of spatial confinement.
The {\it enhanced} electron spin-splitting $U_{o\sigma}$
can be considered as the sum of two terms, $\alpha$ and $\beta$.
$\alpha$ is proportional to the cyclotron gap, $\hbar \omega_c$, while
$\beta$ arises from the exchange interaction between the itinerant carrier
(conduction electron in Ref.~\cite{simserides-prb-2004})
and the localized spins (Mn$^{+2}$ cations in Ref.~\cite{simserides-prb-2004}).
Notice that in such an approximation
the {\it direct} exchange interaction between
the neighboring localized impurity spins is neglected,
being much smaller than the interaction
between impurity spins and carrier spins \cite{brey-guinea:00},
although according to a recent report
it might influence the carrier spin polarization \cite{matsunaka et al}.
The very low $T$ impelled us to a drastic first approximation \cite{simserides-prb-2004},
i.e. to take into account only $\beta$,
and moreover to approximate the corresponding Brillouin function by 1.

\begin{figure}[h!]
\includegraphics[height=6.0cm]{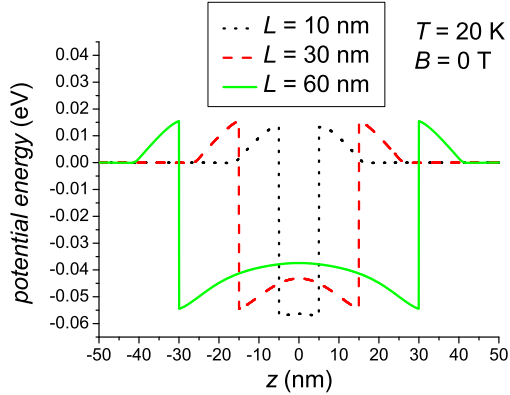}
\caption{(Color online) Sketch of the QW profiles for $T = $ 20 K, $B = $ 0 T
and $N_s =$ 1.566 $\times$ 10$^{11}$ cm$^{-2}$,
for QW widths 10 nm, 30 nm and 60 nm.}
\label{QWs}
\end{figure}

In the present article we attempt a major improvement.
Namely, we examine the relative influence of $\alpha$ and $\beta$
in a wide temperature band (0 to 400 K) and
in a wide in-plane magnetic field band (0 to 20 T), as well as
in a wide range of material parameters,
not necessarily restricting ourselves in the conduction band \cite{vb}.
Our purpose is to systematically study
the influence of the DOS modification on
the spin-subband populations and the spin-polarization of
quasi two-dimensional carriers,
as functions of the in-plane magnetic field and the temperature.
Besides, we indicate the ranges
where the system is completely spin-polarized.
In Section \ref{sec:theory} we introduce our theoretical framework.
In Section \ref{sec:results_and_discussion} we examine
the spin-subband populations and
the spin polarization, $\zeta$, of
non-magnetic-semiconductor (NMS) /
narrow to wide dilute-magnetic-semiconductor (DMS) /
NMS quantum wells (QWs),
as a function of the temperature, $T$,
and the in-plane magnetic field, $B$.
We notice that in the present system due to the influence of carriers,
increase of the QW width transforms the heterostructure
from an ``almost perfect square QW''
to a ``double QW with a soft barrier''
(``a system of two separated heterojunctions'') \cite{QWs_profile}.
Thus, the present heterostructure allows us to study ``single''
as well as ``double'' QWs.
To facilitate the reader, we provide in Fig.~\ref{QWs} sketches of
the self-consistent QW profiles for $T = $ 20 K, $B = $ 0 T
and sheet carrier concentration $N_s =$ 1.566 $\times$ 10$^{11}$ cm$^{-2}$,
for QW widths 10 nm, 30 nm and 60 nm.
We examine how the DOS modification affects $\zeta$
for a wide range of material and structural parameters
focusing on
the quantum well width,
the magnitude of the spin-spin exchange interaction coupling strength,
and the sheet carrier concentration.
Finally, in Section \ref{sec:conclusion} we briefly state our conclusions.

%%%%%%%%%%%%%%%%%%%%%%%%%%%%%%%%%%%%%%%%%%%%%%%%%%%%%%%%%%%%%%%%%%%%%%%%%%%%%%%
\section{Theory}%%%%%%%%%%%%%%%%%%%%%%%%%%%%%%%%%%%%%%%%%%%%%%%%%%%%%%%%%%%%%%%
\label{sec:theory}%%%%%%%%%%%%%%%%%%%%%%%%%%%%%%%%%%%%%%%%%%%%%%%%%%%%%%%%%%%%%
%%%%%%%%%%%%%%%%%%%%%%%%%%%%%%%%%%%%%%%%%%%%%%%%%%%%%%%%%%%%%%%%%%%%%%%%%%%%%%%

Under a magnetic field, $B$, applied parallel to the interfaces,
the equal energy surfaces are gradually distorted.
The density of states deviates from
the ideal step-like form both quantitatively and qualitatively
\cite{simserides-prb-2004}, i.e. it takes the form:

\begin{equation}\label{dos}
\rho({\mathcal E}) = \frac {A \sqrt{2m^*}}{4 \pi^2 \hbar}
\sum_{i,\sigma} \int_{-\infty}^{+\infty} \! dk_x
\frac{\Theta({\mathcal E}-E_{i,\sigma}(k_x))}
{ \sqrt {{\mathcal E}-E_{i,\sigma}(k_x)} },
\end{equation}

\noindent where it is implied that the QW is along the $z$ axis
and the magnetic field is applied along the $y$ axis.
$\Theta$ is the step function, $A$ is the $xy$ area
of the structure, $m^*$ is the effective mass \cite{mstar}.
$E_{i,\sigma}(k_x)$ are the spin-dependent $xz$-plane eigenenergies.
Generally, $E_{i,\sigma}(k_x)$ must be self-consistently calculated
\cite{ees,sj,simserides:99,magnetotransport,PL,simserides-prb-2004}.
Equation~(\ref{dos}) is valid for any type of interplay
between spatial and magnetic confinement.
The $k_x$ dependence in Eq.~(\ref{dos}) increases
the numerical cost by a factor of $10^2-10^3$ in many cases.
This $k_x$ dependence is quite often ``conveniently'' ignored,
although this is only justified for narrow QWs.
However, with the existing computing power,
such a ``simplification'' is not any more necessary.
Only in the limit $B \to 0$, the DOS retains
the {\it occasionally stereotypic} staircase shape
with the {\it ideal} step
$\frac {1}{2} \frac {m^* A}{\pi\hbar^2}$ for each spin.
The opposite asymptotic limit of Eq.~(\ref{dos}) is that of
a simple saddle point, where the DOS diverges logarithmically \cite{lyo:94}.
The DOS modification significantly affects the physical properties
\cite{ees,sj,lyo:94,simserides:99,magnetotransport,PL,simserides-prb-2004}.
For completeness, we notice that Eq.~(\ref{dos}) ignores the effect of disorder
which -with the current epitaxial techniques-
is important when the concentration of magnetic ions is high \cite{matsukura1,matsukura2}.
Disorder will certainly induce some broadening of the spin-subbands.

In DMS structures,
the electron spin-splitting,
$U_{o\sigma}$, is not proportional to the cyclotron gap, $\hbar \omega_c$,
i.e. it acquires the form
\cite{hong:00,ljm:00,kim:02}:

\begin{equation}\label{spin-splitting}
U_{o\sigma} =
\frac {g^*m^*}{2m_e} \hbar \omega_c -
y N_0 J_{sp-d} S B_{S}(\xi) = \alpha + \beta .
\end{equation}

\noindent $\alpha = \alpha (B)$
describes the Zeeman coupling between
the spin of the itinerant carrier and
the magnetic field,
while $\beta = \beta (B,T)$ expresses
the exchange interaction between
the spins of the Mn$^{+2}$ cations and
the spin of the itinerant carrier
(initially supposed to be an electron).
$g^*$ is the g-factor \cite{gstar} of the itinerant carrier.
$y$ is the molecular fraction of Mn.
$N_0$ is the three-dimensional (volume) concentration of cations.
$J_{sp-d}$ is the coupling strength due to the spin-spin exchange interaction
between the d electrons of the Mn$^{+2}$ cations
and the s- or p-band electrons,
and it is negative for conduction band electrons.
The factor $S B_{S}(\xi)$ represents the spin polarization of the Mn$^{+2}$ cations.
The spin of the Mn$^{+2}$ cation is $S =$ 5/2.
$B_{S}(\xi)$ is the standard Brillouin function,
while \cite{ljm:00,brey-guinea:00}

\begin{equation}\label{xi}
\xi=\frac{g_{Mn}\mu_BSB
-J_{sp-d}S \frac{n_{down}-n_{up}}{2}}{k_BT}.
\end{equation}

\noindent $k_B$ is the Boltzmann constant. $\mu_B$ is the Bohr magneton.
$g_{Mn} $ is the $g$ factor of Mn \cite{gmn}.
$n_{down}$ and $n_{up}$ are
the spin-down and spin-up three-dimensional (volume) concentrations
measured e.g. in cm$^{-3}$,
while $N_{s,down}$ and $N_{s,up}$ used below are
the spin-down and spin-up two-dimensional (sheet) concentrations
measured e.g. in cm$^{-2}$.
In Eq.~\ref{xi} (and only there) we approximate
$n_{down}-n_{up} \approx (N_{s,down} - N_{s,up}) / L$,
where $L$ is the QW width.
The first term in the numerator of Eq.~\ref{xi} represents
the contribution of the Zeeman coupling between the localized spin
and the magnetic field.
The second term in the numerator of Eq.~\ref{xi}
(sometimes called ``feedback mechanism'') represents
the kinetic exchange contribution
which -in principle- can induce spontaneous spin-polarization
i.e. in the absence of an external magnetic field \cite{ljm:00}.
Notice that $n_{down} - n_{up}$
is positive for conduction band electrons.
Finally, for conduction band electrons,
the spin polarization can be defined by

\begin{equation}\label{zeta}
\zeta = \frac {N_{s,down}-N_{s,up}}{N_{s}}.
\end{equation}

\noindent $N_{s} = N_{s,down} + N_{s,up}$ is the free carrier
two-dimensional (sheet) concentration.

The use of such a simplified Brillouin-function approach
is quite common when dealing with
quasi two-dimensional systems \cite{teran,koenig_macdonald,hong:00,ljm:00,kim:02}.
This way, the spin-orbit coupling is not taken into account.
This is certainly a simplification, since increasing temperature,
the magnetization of the magnetic ions competes with spin-orbit coupling.
The spin-orbit coupling \cite{matsukura1,matsukura2}
induces temperature dependent spin relaxation.
Therefore, the carriers' spin-polarization does not only depend on
the magnetic order of the magnetic ions,
expressed here with the help of the Brillouin function and
the carriers' spin relaxation influences
the magnetic order of the localized magnetic moments.

The variation of the temperature, $T$,
affects the spin polarization.
The spin polarization is also influenced by the magnetic field,
in an opposite manner i.e.
$B$ tends to align the spins.
Furthermore, for each type of spin population,
the in-plane magnetic field
-via the distortion of the DOS-
redistributes the electrons between the subbands.
Consequently, the spin polarization can be tuned
by varying the temperature and the magnetic field.
Indeed, preliminary conduction band calculations for specific values
of the material parameters,
for very narrow quantum wells,
have shown \cite{simserides_conf} that
when the ``feedback mechanism''
due to the difference between the populations of
the spin down and the spin up electrons can be neglected,
the spin polarization vanishes for $B \to 0$.
The analysis presented above can be useful for p-doped structures,
assuming -as usual- that a single valence band description is
a fair first approximation.

%%%%%%%%%%%%%%%%%%%%%%%%%%%%%%%%%%%%%%%%%%%%%%%%%%%%%%%%%%%%%%%%%%%%%%%%%%%%%%%
\section{Results and discussion}%%%%%%%%%%%%%%%%%%%%%%%%%%%%%%%%%%%%%%%%%%%%%%%
\label{sec:results_and_discussion}%%%%%%%%%%%%%%%%%%%%%%%%%%%%%%%%%%%%%%%%%%%%%
%%%%%%%%%%%%%%%%%%%%%%%%%%%%%%%%%%%%%%%%%%%%%%%%%%%%%%%%%%%%%%%%%%%%%%%%%%%%%%%
Initially we consider heterostructures of the type
n-doped ZnSe / Zn$_{1-x-y}$Cd$_x$Mn$_y$Se /n-doped ZnSe.
Let us take $y = 0.035$,
$- y N_0 J_{sp-d} =$ 0.13 Hartree$^*$, and
the conduction band offset, $\Delta U_{cb} =$ 1 Hartree$^*$ \cite{hong:00}.
We notice that for ZnSe, 1 Hartree$^* \approx$ 70.5 meV.
ZnSe has a sphalerite-type structure and
the lattice constant is $\sim$ 0.567 nm.
Hence, $- J_{sp-d} \approx 12 \times 10^{-3}$ eV nm$^3$.
This is one order of magnitude smaller than
the value commonly used for the III-V Ga(Mn)As valence band system
($J_{pd} = 15 \times 10^{-2}$ eV nm$^3$)
\cite{brey-guinea:00,ljm:00,kim:02}.

%% FIGURE
\begin{figure}[h!]
\includegraphics[height=4.3cm]{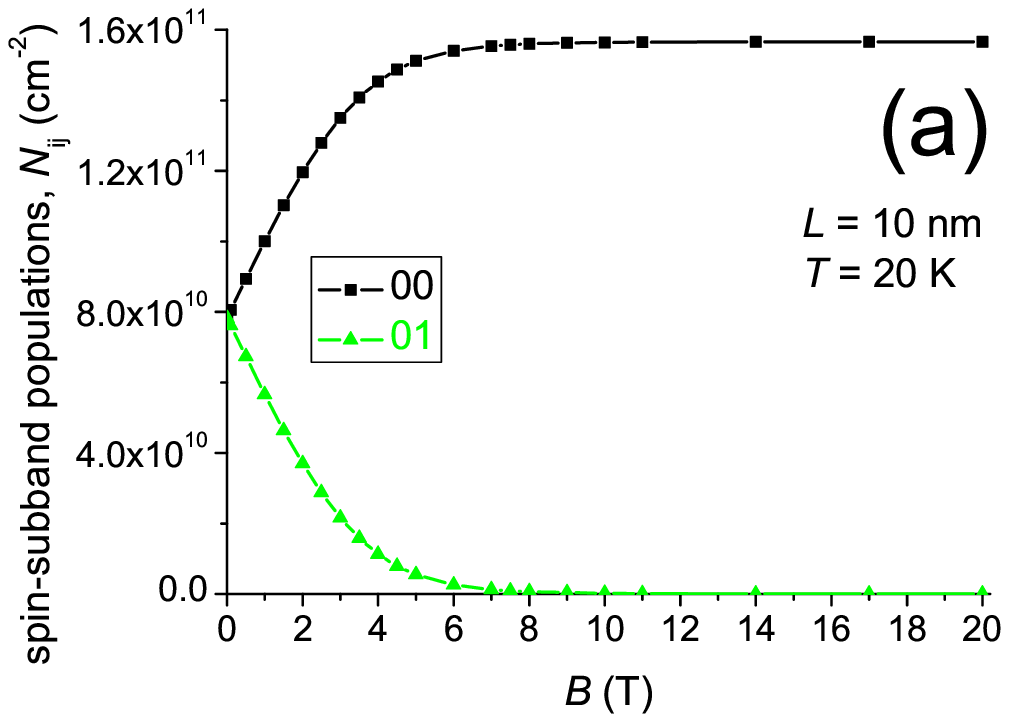}
\includegraphics[height=4.3cm]{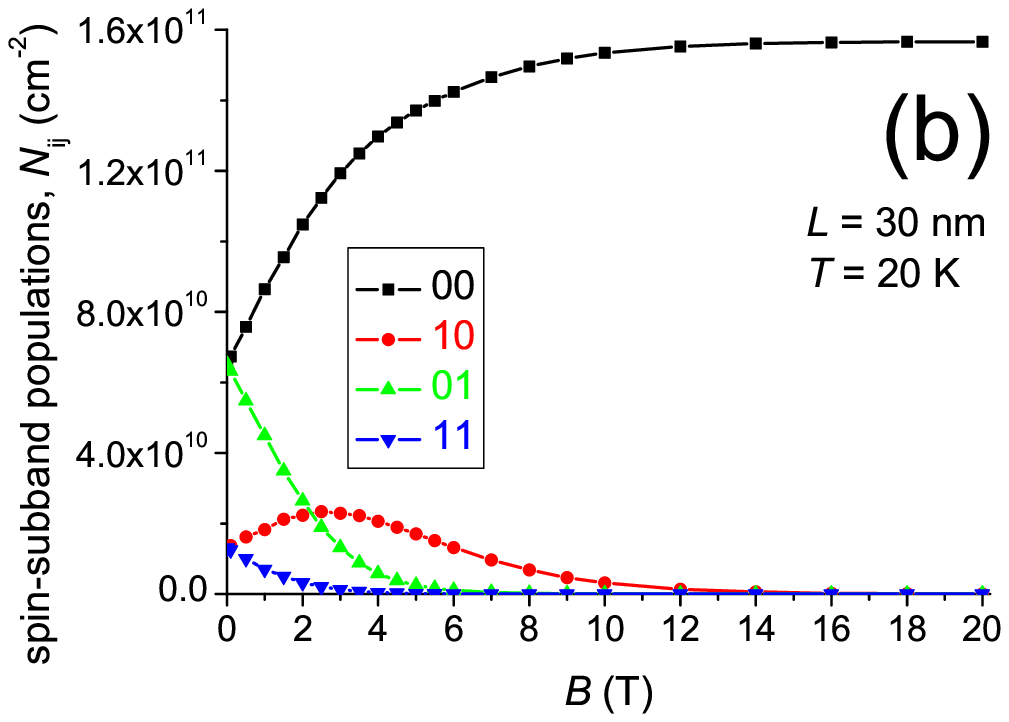}
\includegraphics[height=4.3cm]{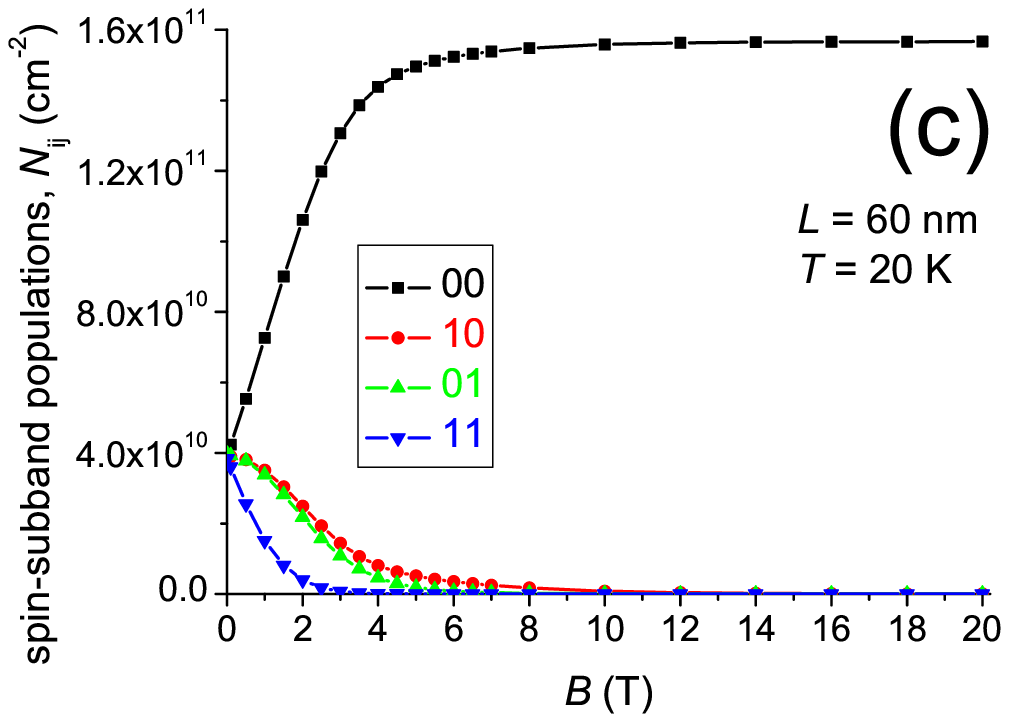}
\includegraphics[height=4.3cm]{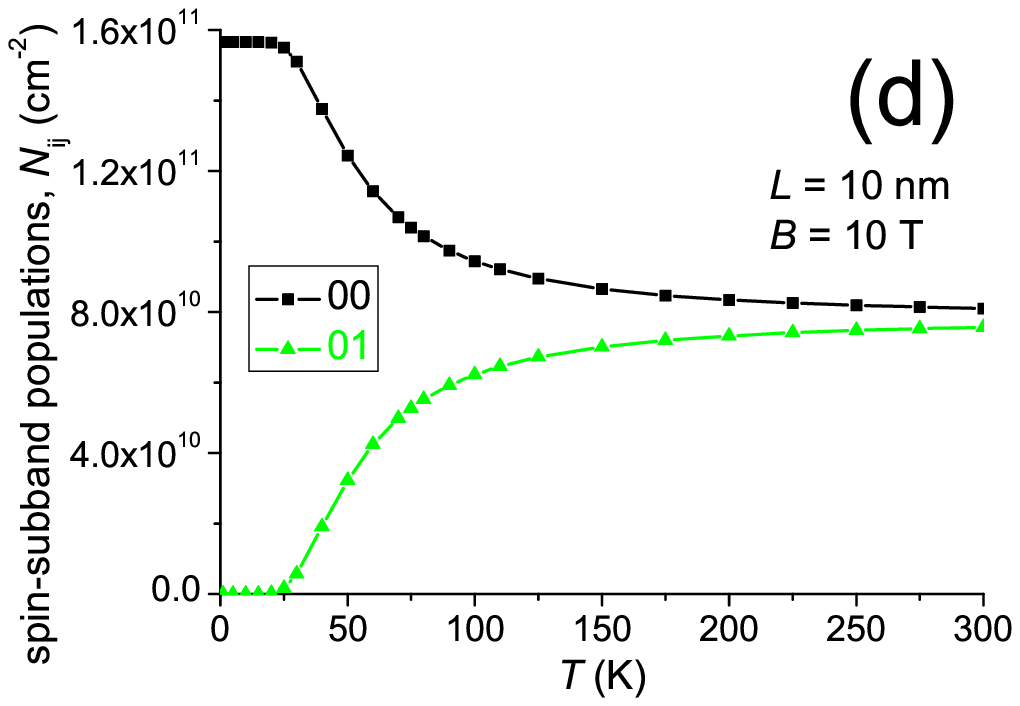}
\includegraphics[height=4.3cm]{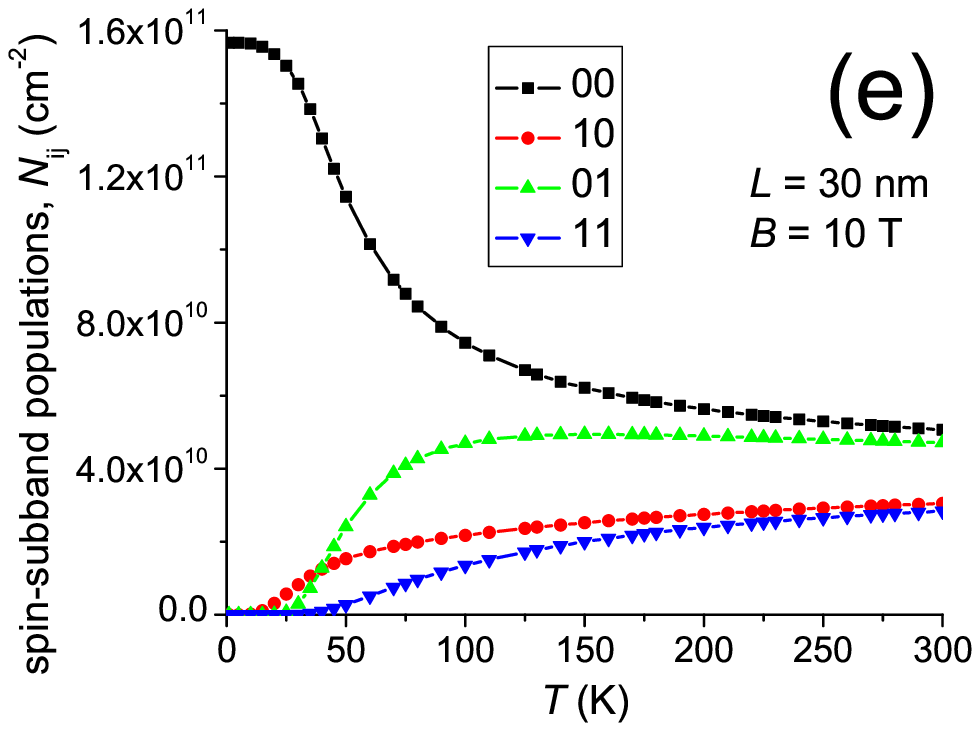}
\includegraphics[height=4.3cm]{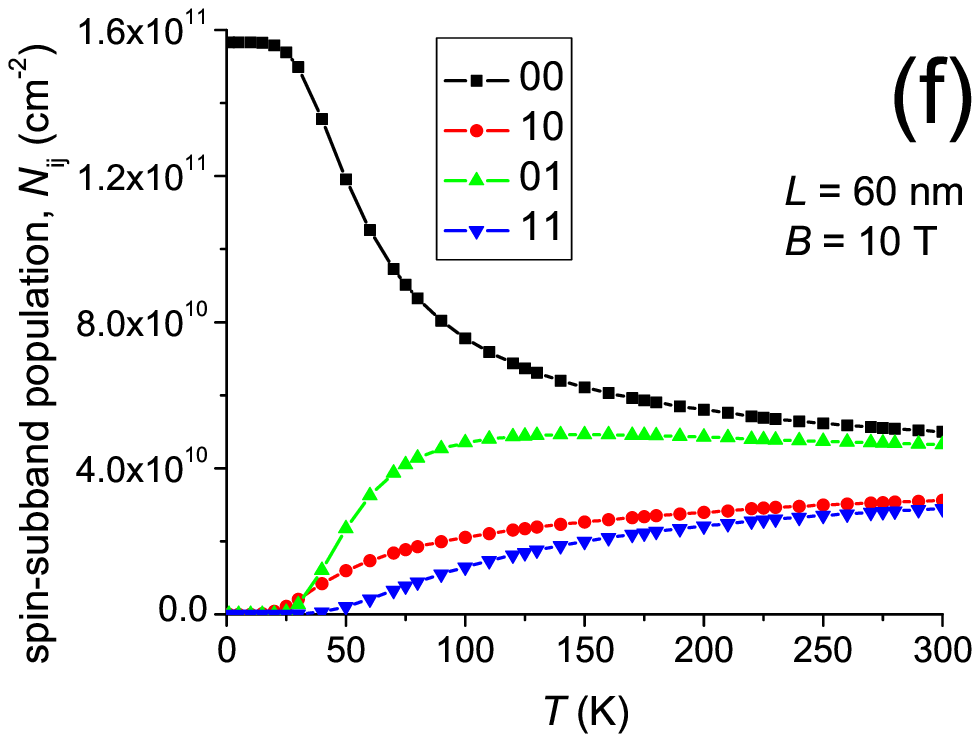}
\caption{(Color online) The spin-subband populations
$N_{ij}=N_{ij}(B)$ for $T = $ 20 K (first row), and
$N_{ij}=N_{ij}(T)$ for $B = $ 10 T (second row)
of a n-doped ZnSe / Zn$_{1-x-y}$Cd$_x$Mn$_y$Se /n-doped ZnSe QW with $y = 0.035$.
$- J_{sp-d} = 12 \times 10^{-3}$ eV nm$^3$.
00 stands for the ground-state spin-down-subband,
10        for the 1st excited spin-down-subband,
01        for the ground-state spin-up-subband, and
11 represents the 1st excited spin-up-subband.
Each column corresponds to different well width i.e.
$L = $ 10 nm, 30 nm, and 60 nm.}
\label{fig:NijBT_10nm_30nm_60nm}
\end{figure}

Figure~\ref{fig:NijBT_10nm_30nm_60nm} depicts
the spin-subband populations, $N_{ij}$ as a function of $B$ (a-c),
and as a function of $T$ (d-f),
for three different well widths, namely
(a,d) for $ L = $ 10 nm
(b,e) for $ L = $ 30 nm, and
(c,f) for $ L = $ 60 nm.
Initially, we deliberately keep
the total sheet carrier concentration constant
($N_s =$ 1.566 $\times$ 10$^{11}$ cm$^{-2}$),
assuming that all dopants are ionized.
In (a-c) $ T = $ 20 K.
In (d-f) $ B = $ 10 T.
The pair $ij$ is defined in the following manner:
00 symbolizes the ground-state spin-down-subband,
10            the 1st excited spin-down-subband,
01            the ground-state spin-up-subband, and finally
11 symbolizes the 1st excited spin-up-subband.
Due to the small value of $J_{sp-d}$,
the influence of the ``feed-back mechanism''
due to the difference between spin-down and spin-up concentrations
is negligible in the present system.
Indeed, since $-J_{sp-d} \frac{n_{down}-n_{up}}{2}$ is negligible here,
then for $B =$ 0, it follows that
(a) $\xi \approx$ 0, thus  $B_{S}(\xi) \approx$ 0, therefore $\beta \approx$ 0, and
(b) $g^* \mu_B B = \alpha =$ 0.
Hence, $U_{o\sigma} \approx $ 0, and consequently $\zeta \approx $ 0.
In fact, inspection of Figs.\ref{fig:NijBT_10nm_30nm_60nm} (a-c),
reveals that for $B =$ 0,
in Fig.\ref{fig:NijBT_10nm_30nm_60nm}(a) $N_{00} = N_{01}$,
in Fig.\ref{fig:NijBT_10nm_30nm_60nm}(b) $N_{00} = N_{01}$ and $N_{10} = N_{11}$, and
in Fig.\ref{fig:NijBT_10nm_30nm_60nm}(c) $N_{00} = N_{01}$ and $N_{10} = N_{11}$.
For the very wide quantum well ($ L = $ 60 nm),
as expected \cite{QWs_profile},
the four spin-subbands are almost equally populated for $ B = $ 0.
Increasing $B$, we observe that there are two mechanisms which cause depopulations:
(I) The increase of $U_{o\sigma}$ eliminates spin-up electrons, namely
$N_{01}$ and $N_{11}$ continuously decrease, increasing $B$.
(II) The DOS modification which depopulates all excited states,
regardless of their spin \cite{simserides:99,simserides-prb-2004}, namely
the eventual decay of $N_{10}$.
Finally, in Figs.\ref{fig:NijBT_10nm_30nm_60nm} (d-f),
we witness the survival of only $N_{00}$ at very low $T$,
since $U_{o\sigma}$ acquires its bigger value at zero temperature.
Increasing $T$, $U_{o\sigma}$ decreases,
augmenting the influence of the spin-up electrons.

\begin{figure}[h!]
\includegraphics[height=6.0cm]{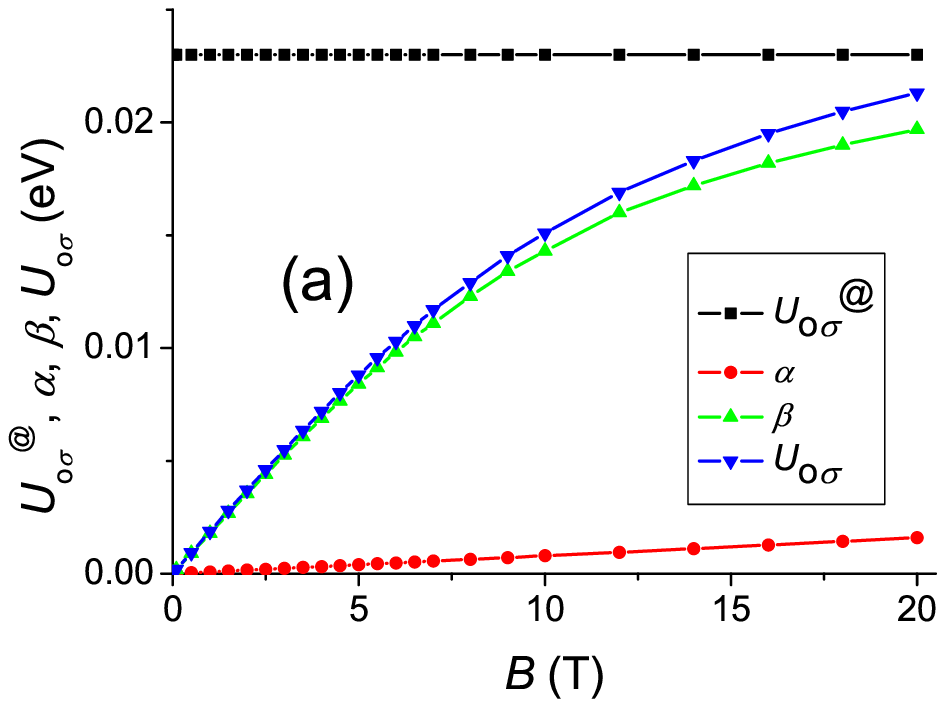}
\includegraphics[height=6.0cm]{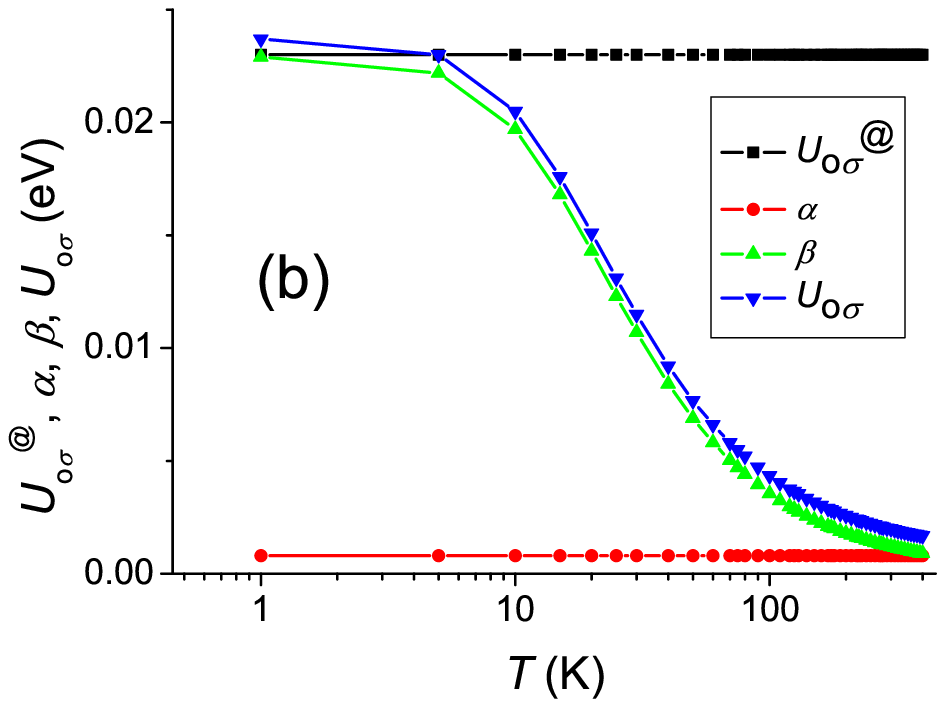}
\caption{(Color online) The relative influence of
the Zeeman term, $\alpha$, and
the exchange term $\beta$
in wide $B$ and $T$ ranges,
for a n-doped ZnSe / Zn$_{1-x-y}$Cd$_x$Mn$_y$Se /n-doped ZnSe QW
with $L =$ 60 nm and $y = 0.035$.
$- J_{sp-d} = 12 \times 10^{-3}$ eV nm$^3$.
Of course, $\alpha =\alpha (B)$, while $\beta = \beta (B,T)$.
(a) $\alpha=\alpha(B)$,          $\beta=\beta(B)$ for $T = $ 20 K.
(b) $\alpha=\alpha(T)=constant$, $\beta=\beta(T)$ for $B = $ 10 T.
$L = $ 60 nm.
Each panel also contains the spin-splitting,
$U_{o\sigma} = \alpha + \beta$,
as well as the value of
of the spin-splitting used
in our previous low-$T$ calculation \cite{simserides-prb-2004}
(i.e. taking into account only $\beta$ and
approximating the corresponding Brillouin function by 1),
$U_{o\sigma}^{@}$.
For comparison we notice that the conduction band-offset,
$\Delta U_{cb}$ = 1 Hartree$^* \approx$ 70.5 meV.
}
\label{fig:alpha-beta-BT_60nm}
\end{figure}

Figure \ref{fig:alpha-beta-BT_60nm} depicts
the relative influence of the Zeeman term, $\alpha$, and
the exchange term, $\beta$,
in wide $B$ and $T$ ranges,
for a n-doped ZnSe / Zn$_{1-x-y}$Cd$_x$Mn$_y$Se /n-doped ZnSe QW
with $L =$ 60 nm and $y = 0.035$.
In reality $L$ is of no importance here due to the negligible
impact of the ``feedback mechanism'' with these material parameters.
For comparison we notice that the conduction band-offset,
$\Delta U_{cb}$ = 1 Hartree$^* \approx$ 70.5 meV.
The spin splitting in the present article,
$U_{o\sigma} = \alpha + \beta$,
while $U_{o\sigma}^{@}$ was used
in our previous low-$T$ calculations \cite{simserides-prb-2004}
($B_{5/2}(\xi)$ approximated by 1, and $\alpha$ ignored).
Figure \ref{fig:alpha-beta-BT_60nm} elaborates
the competition between $B$ (aligning spins) and $T$ (bringing on anarchy).
Figure \ref{fig:alpha-beta-BT_60nm}b
justifies our previous low-temperature approximation:
at low enough $T$, $U_{o\sigma} \approx U_{o\sigma}^{@}$.
At higher temperatures, $B_{5/2}(\xi)$ cannot be approximated with 1.
As $k_BT$ increases, $\xi$ decreases, and consequently
$B_{5/2}(\xi) <$ 1.
In other words, increasing $T$,
the spin-splitting decreases
allowing enhanced contribution of the spin-up electrons
to the system's properties.
Finally we notice that an opposite sign of $g^*$
(e.g. CdTe vs. ZnSe) is expected to have small effect on the results
since the most important term is $\beta$.

\begin{figure}
\includegraphics[height=6.0cm]{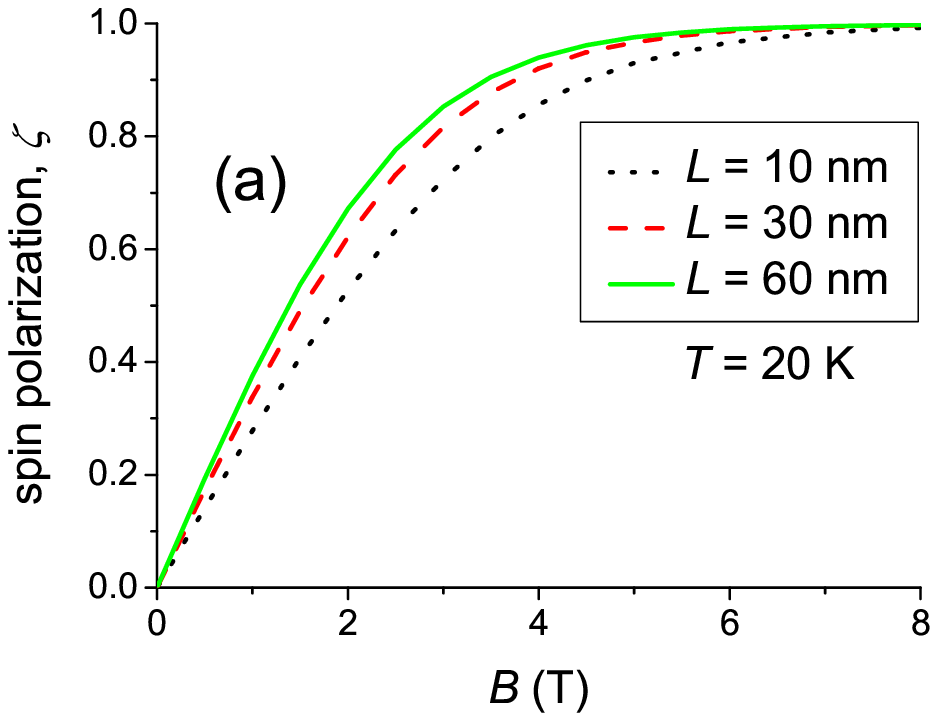}
\includegraphics[height=6.0cm]{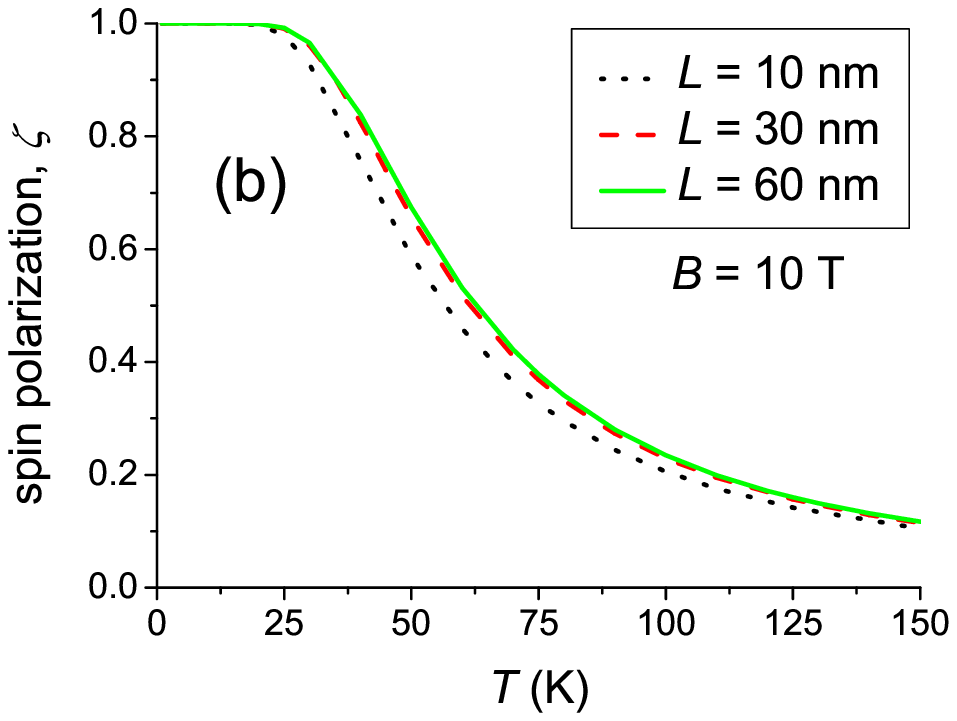}
\caption{(Color online) The spin polarization, $\zeta$, tuned by varying:
(a) the in-plane magnetic field, $B$, keeping $T = $ 20 K (left panel), and
(b) the temperature, $T$, keeping $B =$ 10 T (right panel),
for different well widths, $L = $ 10 nm, 30 nm, and 60 nm.
$- J_{sp-d} = 12 \times 10^{-3}$ eV nm$^3$. }
\label{zeta_of_B_and_T_10nm_30nm_60nm}
\end{figure}

Figure \ref{zeta_of_B_and_T_10nm_30nm_60nm} depicts
the spin polarization tuned by varying
the parallel magnetic field and the temperature,
for different choices of the well width.
Since for $B \geq $ 8 T, $\zeta = $ 1,
only the range $B \in$ [0, 8 T] is presented in
Fig.~\ref{zeta_of_B_and_T_10nm_30nm_60nm}a.
Since for $T \geq $ 150 K, $\zeta$ is less than $\approx$ 0.1,
only the range $T \in$ [0, 150 K] is presented in
Fig.~\ref{zeta_of_B_and_T_10nm_30nm_60nm}b.
Because of the DOS modification \cite{simserides-prb-2004},
resulting in different distribution of electrons
among the spin-subbands (cf. Fig.\ref{fig:NijBT_10nm_30nm_60nm}),
we observe a clear dependence of $\zeta = \zeta (L)$, i.e.
$\zeta (L = $ 60 nm) $ >  \zeta ( L = $ 30 nm) $ > \zeta ( L = $ 10 nm).
We also observe that for $B =$ 0, $\zeta$ vanishes,
i.e. there is no spontaneous spin polarization phase
due to the tiny ``feedback mechanism''
for this choice of material parameters.

\begin{figure}
\includegraphics[height=6.0cm]{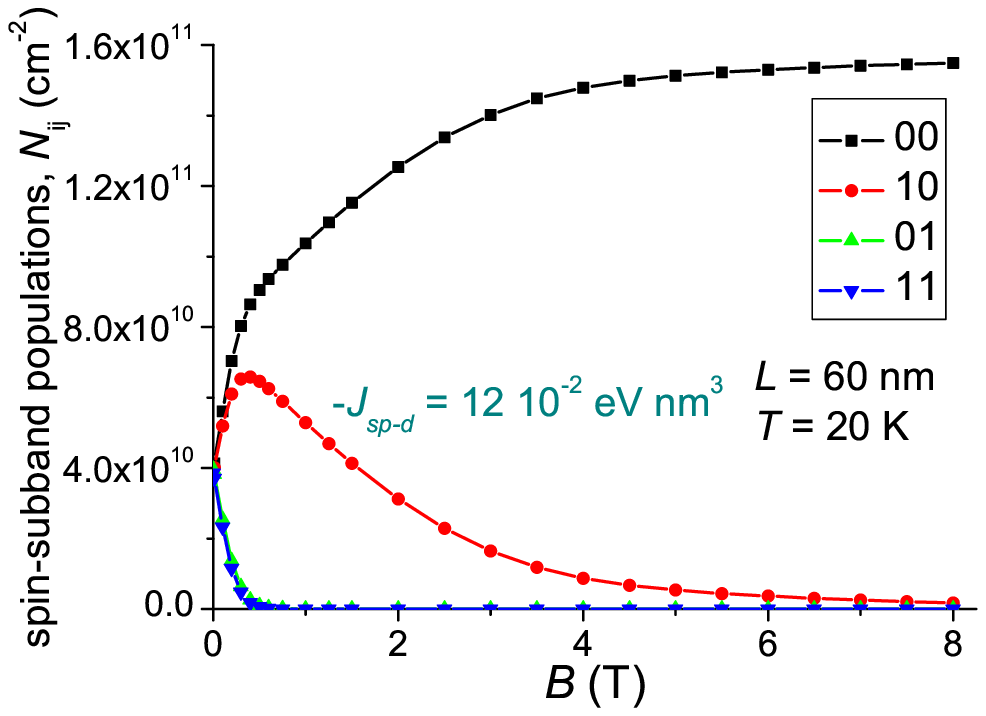}
\includegraphics[height=6.0cm]{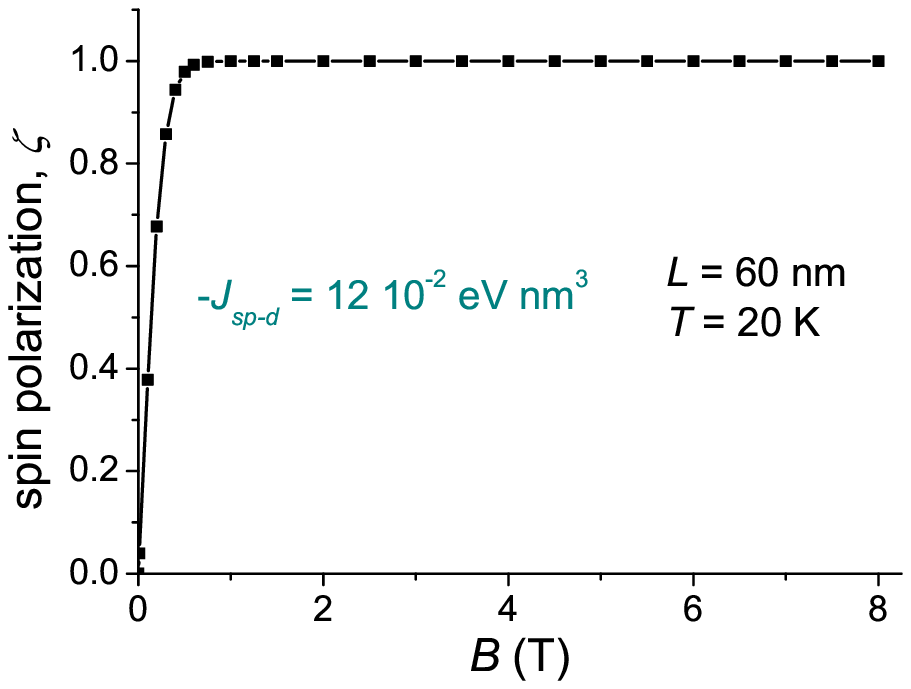}
\caption{(Color online)
The spin-subband populations, $N_{ij}$ (left panel) and
the spin polarization, $\zeta$ (right panel)
tuned by varying $B$
for $L = $ 60 nm, $T = $ 20 K,
using $- J_{sp-d} = 12 \times 10^{-2}$ eV nm$^3$.}
\label{ssp_and_zeta_of_B_Fact}
\end{figure}

Subsequently we deliberately increase $- J_{sp-d}$ by an order of magnitude
i.e. we present in Fig.~\ref{ssp_and_zeta_of_B_Fact} results with
$- J_{sp-d} = 12 \times 10^{-2}$ eV nm$^3$
(which is of a little smaller magnitude than
the value commonly used \cite{brey-guinea:00,ljm:00,kim:02}
for the III-V Ga(Mn)As valence band system,
$J_{pd} = 15 \times 10^{-2}$ eV nm$^3$).
$L = $ 60 nm and $T = $ 20 K.
Comparing
Fig.~\ref{ssp_and_zeta_of_B_Fact} with
Fig.~\ref{fig:NijBT_10nm_30nm_60nm}c and
Fig.~\ref{zeta_of_B_and_T_10nm_30nm_60nm}a
we observe that:
($\alpha '$) The greater value of $- J_{sp-d}$ makes it much easier to attain
a completely spin-polarized system ($\zeta =$ 1)
i.e. for $B \geq $ 1 T instead of $B \geq $ 8 T.
($\beta '$) Initially, increasing $B$, due to the increased $U_{o\sigma}$,
$N_{10}$ grows, in contrast to Fig.~\ref{fig:NijBT_10nm_30nm_60nm}c.
Naturally, subsequently $N_{10}$ is depopulated because of
the in-plane magnetic field induced DOS modification.
($\gamma '$) Although the system is more susceptible to spin-polarization,
still, practically no spontaneous spin-polarization phase exists for $B =$ 0,
at this temperature.

\begin{figure}
\includegraphics[height=6.0cm]{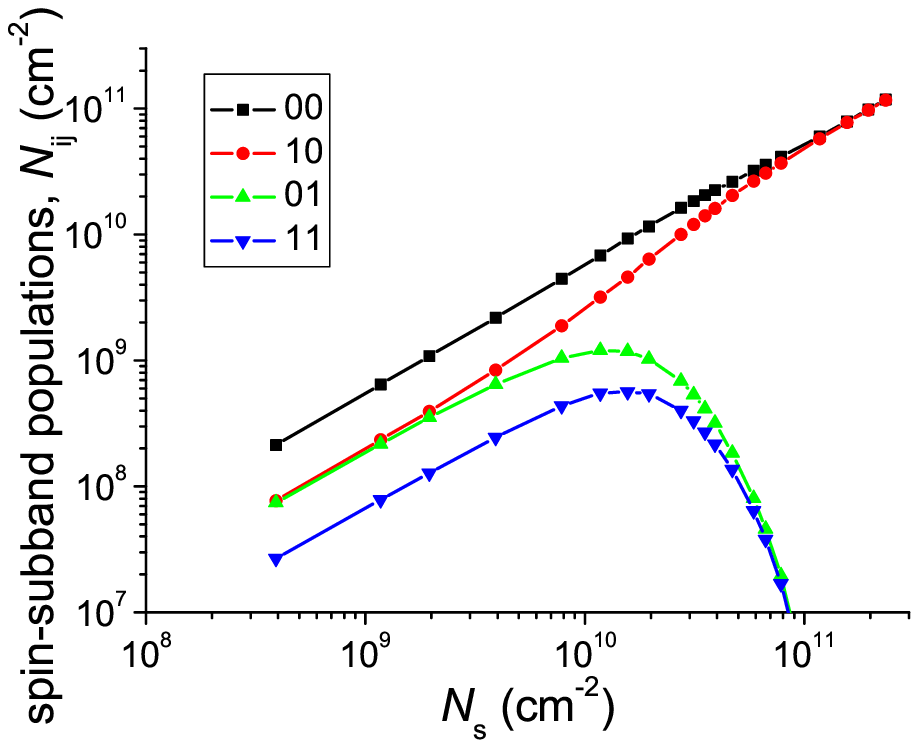}
\includegraphics[height=6.0cm]{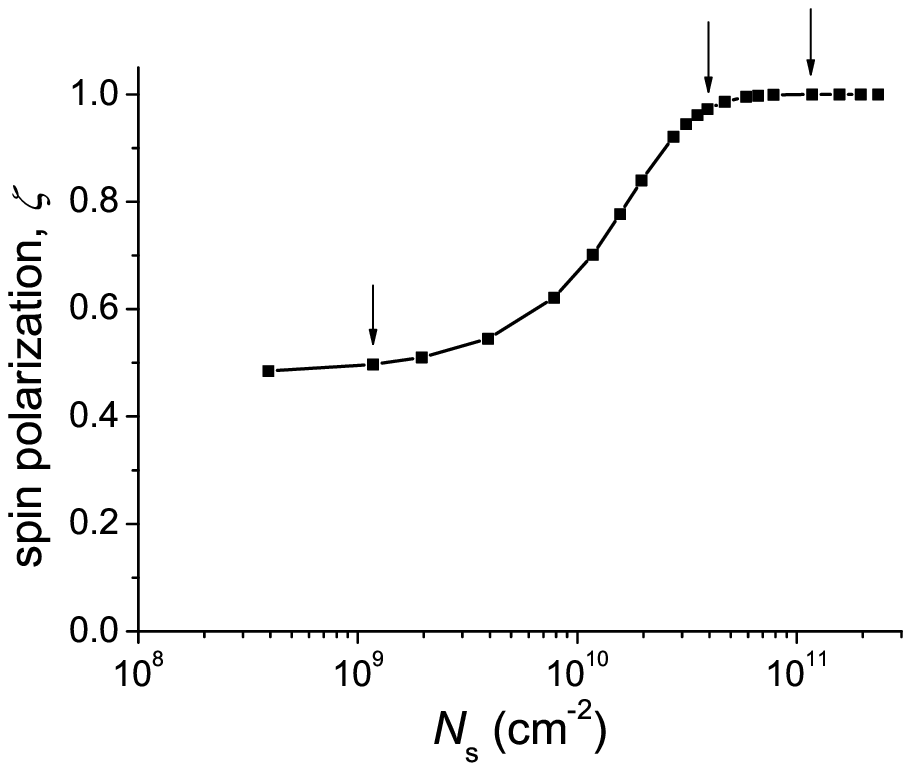}
\caption{(Color online)
The spin-subband populations, $N_{ij}$ (left panel) and
the spin polarization, $\zeta$ (right panel),
tuned by varying the sheet carrier concentration, $N_s$,
for $L = $ 60 nm, $T = $ 20 K and $B =$ 0.01 T,
using  $J = 12 \times 10^{-1}$ eV nm$^3$.
The little arrows indicate $N_s$ values
where we also compare with $B =$ 0.0001 T in the text.}
\label{ssp_and_zeta_of_Ns_FF}
\end{figure}

Up to now, we have deliberately kept
the total sheet carrier concentration constant.
Below we examine the influence of $N_s$ on
the spin-subband populations and the spin polarization
for different values of
the magnitude of the spin-spin exchange interaction, $J$.
Since $N_s$ is affected by many factors
(QW profile, material properties,
valence-band- or conduction-band-based structures etc)
we have decided to use $J$ as a parameter here.
Naturally, in a heterostructure where higher $N_s$ can be achieved
we may require smaller values of $J$ in order to completely spin-polarize carriers.
Using the rest material parameters as above but modifying $J$,
we have systematically studied the $N_s$ influence.
For $J = 12 \times 10^{-2}$ eV nm$^3$
there is a very small influence of $N_s$ on $\zeta$.
The situation changes using $J = 12 \times 10^{-1}$ eV nm$^3$.
Figure \ref{ssp_and_zeta_of_Ns_FF} shows
$N_{ij}$ and $\zeta$
tuned by varying $N_s$
for $L = $ 60 nm, $T = $ 20 K and $B =$ 0.01 T,
using  $J = 12 \times 10^{-1}$ eV nm$^3$.
We observe that increase of $N_s$
from $\approx$ 1.0 $\times$ 10$^{9}$ cm$^{-2}$
  to $\approx$ 1.0 $\times$ 10$^{11}$ cm$^{-2}$
is sufficient to completely spin-polarize carriers.
This is purely due to the ``feedback mechanism''
stemming from the difference between the populations of
spin-down and spin-up carriers.
If we decrease $B$ from 0.01 T to 0.0001 T, then e.g.
(a) for $N_s =$ 1.175 $\times$ 10$^{9}$ cm$^{-2}$,
$\zeta$ changes from 0.497 to 0.005,
(b) for $N_s =$ 3.917 $\times$ 10$^{10}$ cm$^{-2}$,
$\zeta$ changes from 0.973 to 0.909,
however,
(c) for $N_s =$ 1.175 $\times$ 10$^{11}$ cm$^{-2}$,
$\zeta$ remains 1.

%%%%%%%%%%%%%%%%%%%%%%%%%%%%%%%%%%%%%%%%%%%%%%%%%%%%%%%%%%%%%%%%%%%%%%%%%%%%%%%
\section{Conclusion}%%%%%%%%%%%%%%%%%%%%%%%%%%%%%%%%%%%%%%%%%%%%%%%%%%%%%%%%%%%%%
\label{sec:conclusion}%%%%%%%%%%%%%%%%%%%%%%%%%%%%%%%%%%%%%%%%%%%%%%%%%%%%%%%%%%%
%%%%%%%%%%%%%%%%%%%%%%%%%%%%%%%%%%%%%%%%%%%%%%%%%%%%%%%%%%%%%%%%%%%%%%%%%%%%%%%
We have studied the spin-subband structure of
quasi two-dimensional carriers
in dilute-magnetic-semiconductor-based heterostructures,
under the influence of
an in-plane magnetic field.
The proper density of states was used for the first time,
incorporating the dependence on the in-plane wave vector
perpendicular to the in-plane magnetic field.
We have examined the interplay between different degrees of
spatial and magnetic confinement,
as well as the influence of temperature in a wide range.
We have systematically studied
the spin-subband populations and the spin-polarization
as functions of the temperature and the in-plane magnetic field.
We have examined a wide range of material and structural parameters,
focusing on the quantum well width,
the magnitude of the spin-spin exchange interaction,
and the sheet carrier concentration.
In particular we have shown that
with sufficient magnitude of the spin-spin exchange interaction,
the sheet carrier concentration emerges as an important factor
to manipulate the spin-polarization,
inducing spontaneous (i.e. for $B \to $ 0) spin-polarization.
We have shown how at low temperatures
the spin-splitting acquires its bigger value and
how it decreases at higher temperatures.
Increasing the in-plane magnetic field, the spin-splitting increases
inducing depopulations of the ``minority''-spin subbands.
Moreover, the DOS modification induces depopulations of
all energetically higher subbands.
Finally, we have indicated the ranges where the system is completely spin-polarized. \\

%\end{multicols}


\begin{thebibliography}{33}
\bibitem[11]{gstar} For conduction band electrons,
we use $g^* = $ 1.37  for ZnSe \cite{venghaus},
and   $g^* = $ -1.644 for CdTe \cite{sirenko}.
Some of the values found in the literature are quoted below:

\begin{tabular}{|l|r|r|r|r|}
\hline
$g^*$ & Ref.\cite{venghaus} & Ref.\cite{sirenko} & Ref.\cite{willatzen} & Ref.\cite{cardona} \\
\hline
ZnSe &  1.37 &         & 1.15  &  1.7 \\
\hline
CdTe &       &  -1.644 &-1.59  & -1.1 \\
\hline
\end{tabular}
\bibitem[21]{vb} Strictly speaking, the analysis presented here
is only valid for the conduction band.
Things are, as usual, more complicated in the valence band \cite{winkler}.
However, our analysis might be helpful for comprehending
valence-band-based QWs under parallel magnetic field,
assuming that a single-valence-band description is a reasonable compromise.
\bibitem[24]{mstar} For ZnSe we use $m^* = 0.16 m_e$ \cite{merz}.
$m_e$ is the electron mass.
\end{thebibliography}
\end{document}